\documentclass[copyright,creativecommons]{eptcs}
\providecommand{\event}{CompMod 2009}
\providecommand{\volume}{??}
\providecommand{\anno}{2009}
\providecommand{\firstpage}{1}
\usepackage{breakurl}
\usepackage{graphics}
\usepackage{epsfig}
\usepackage{amsmath}
\usepackage{amssymb}
\usepackage{theorem}

\title{BioDiVinE: A Framework for Parallel Analysis of Biological Models}
\author{J.~Barnat\thanks{J.~Barnat has been supported by the Czech Science Foundation grant No. 201/09/P497. }, L.~Brim\thanks{L.~Brim has been supported by the Grant Agency of Czech Republic grant No. 201/09/1389.}, I.~\v{C}ern\'a, S.~Dra\v{z}an\thanks{S.~Dra\v{z}an and J.~Fabrikov\'a have been supported by the Academy of Sciences of CR grant No.~1ET408050503.}, J.~Fabrikov\'a, J.~L\'an\'{\i}k, and D. \v{S}afr\'anek
\institute{Faculty of Informatics\\
Masaryk University\\
Botanick\'a 68a, Brno, Czech Republic}
\email{safranek@fi.muni.cz}
\and
H.~Ma\thanks{I.~\v{C}ern\'a, D.~\v{S}afr\'anek, and H.~Ma have been supported by the FP6 project No.~NEST-043235 (EC-MOAN).}
\institute{Informatics Life-Sciences Institute\\
School of Informatics, University of Edinburgh\\
11 Crichton Street, Edinburgh, EH8 9LE, United Kingdom}
}
\def\titlerunning{BioDiVinE}
\def\authorrunning{J.~Barnat et.al.}

\newcommand{\biodve}{\texttt{BioDiVinE {}}}
\newcommand{\G}{\mathbf{G}}
\newcommand{\F}{\mathbf{F}}
\newcommand{\la}[1]{\stackrel{#1}{\leftarrow}}
\newcommand{\ra}[1]{\stackrel{#1}{\rightarrow}}

\newtheorem{theorem}{Theorem}[section]

\begin{document}
\maketitle

\begin{abstract}
  In this paper a novel tool \biodve for parallel analysis of
  biological models is presented. The tool allows analysis of
  biological models specified in terms of a set of chemical reactions.
  Chemical reactions are transformed into a system of multi-affine
  differential equations. \biodve employs techniques for finite
  discrete abstraction of the continuous state space. At that level,
  parallel analysis algorithms based on model checking are provided.
  In the paper, the key tool features are described and their application is
  demonstrated by means of a case study.
\end{abstract}

\section{Introduction}



The central interest of systems biology is investigation of the
response of the organism to environmental events (extra-cellular or
intra-cellular signals). Even in procaryotic organisms, a single
environment event causes a response induced by the interaction of
several interwoven modules with complex dynamic behaviour, acting on rapidly
different time scales. In higher organisms, these modules form large
and complex interaction networks.  For instance, a human cell contains
in the order of 10,000 substances which are involved in 15,000
different types of reactions.  This gives rise to a giant interaction
network with complex positive and negative regulatory feedback loops.

In order to deal with the complexity of living systems, experimental methods
have to be supplemented with mathematical modelling and computer-supported
analysis. One of the most critical limitations in applying current approaches
to modelling and analysis is their pure scalability. Large models require
powerful computational methods, the hardware infrastructure is available
(clusters, GRID, multi-core computers), but the parallel (distributed)
algorithms for model analysis are still under development.



In this paper we present the tool \biodve for parallel analysis of
biological models. \biodve considers the model in terms of chemical equations. The main features of the tool are the following:
\begin{itemize}
\item user interface for specification of models in terms of chemical equations
\item formal representation of the model by means of multi-affine ODEs
\item setting initial conditions and parameters of the kinetics
\item setting parameters of the discrete abstraction
\item graphical simulation of the discretized state space
\item model checking analysis.
\end{itemize}

As an abstraction method, \biodve adapts the rectangular abstraction approach of multi-affine ODEs
mathematically introduced in~\cite{BH06} and algorithmically tackled in
\cite{KB08,batt-tacas07} by means of a parallel on-the-fly state space generator. The character of abstraction provided by this discretization technique is conservative with respect to the most dynamic properties that are of interest. However, there is the natural effect of adding false-positive behaviour, in particular, the abstracted state space includes trajectories which are not legal in the original continuous model.

The structure of the paper is the following.
Section~\ref{sec:background} gives a brief overview of the underlying
abstraction technique and the model checking approach.
Section~\ref{sec:biodivine} presents in step-by-step fashion the key
features of current version of BioDiVinE.  Section~\ref{sec:casestudy}
provides a case study of employing BioDiVinE for analysis of a
biological pathway responsible for ammonium transport in bacteria
Escherichia Coli.

\subsection{Related Work}


In our previous work~\cite{TCS} we have dealt with parallel model
checking analysis of piece-wise affine ODE models~\cite{JGHP03}. The
method allows fully qualitative analysis, since the piece-wise affine
approximation of the state space does not require numerical
enumeration of the equations. Therefore that approach, in contrast to
the presented one, is primarily devoted for models with unknown kinetic
parameters. The price for this feature is higher time complexity of
the state space generation. In particular, time appears there more
critical than space while causing the parallel algorithms not to scale
well.

In the current version of BioDiVinE all the kinetic parameters are
required to be numerically specified.  In such a situation there is an
alternative possibility to do LTL model checking directly on numerical
simulations~\cite{Gilbert,CFS06}.  However, in
the case of unknown initial conditions there appears the need to
provide large-scale parameter scans resulting in huge number of
simulations.  On the contrary, the analysis conducted with BioDiVinE
can be naturally generalised to arbitrary intervals of initial
conditions by means of rectangular abstraction.

Rectangular abstraction of dynamic systems has been employed
in~\cite{KB08} for reachability analysis. The method has been
supported by experiments performed on a sequential implementation in
Matlab. The provided experiments have showed that for models with $10$
variables the reachability analysis ran out of memory after $2$ hours
of computation. This gave us the motivation to employ parallel
algorithms. Moreover, we generalise the analysis method to full LTL
model checking.

There is another work that employs rectangular abstraction for dynamic
systems~\cite{BBW08}. The framework is suitable for deterministic
modelling of genetic regulatory networks.  The rectangular abstraction
relies on replacing S-shaped regulatory functions with piece-wise
linear ramp functions. The partitioning of the state space is
determined by parameters of the ramp functions. In contrast to that
work, we consider directly general multi-affine systems with
arbitrarily defined abstraction partitions.

\section{Background}
\label{sec:background}


\subsection{Modelling Approach}

The most widely used approach to modelling a system of biochemical
reactions is the continuous approach of ordinary differential
equations (ODEs).  We consider a special class of ODE systems in the
form $\dot{x} = f(x)$ where $x=(x_1,\ldots,x_n)$ is a vector of
variables and $f=(f_1,...,f_n):\mathbb{R}^n \rightarrow \mathbb{R}^n$
is a vector of multiaffine functions. A multiaffine function is a
polynomial in the variables $x_1,...,x_n$ where the degree of any
variable is at most $1$.  Variables $x_i$ represent continuous
concentrations of species. Multiaffine ODEs can express reactions in
which the stoichiometry coefficients of all reactants are at most $1$.
The system of ODEs can be constructed directly from the stoichiometric
matrix of the biochemical system~\cite{Fe87}.

\begin{figure}[!h]
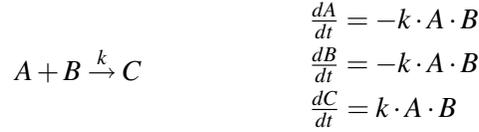

\begin{center}
\begin{minipage}{3cm}
$$A + B \ra{k} C$$
\end{minipage}
\hspace{1cm}
\begin{minipage}{3cm}
$$\begin{array}{l}
\frac{dA}{dt} = -k\cdot A\cdot B\\[1mm]
\frac{dB}{dt} = -k\cdot A\cdot B\\[1mm]
\frac{dC}{dt} = k\cdot A\cdot B
\end{array}
$$
\end{minipage}
\end{center}
\caption{Example of a multi-affine system}
\label{fig:multiaffine}
\end{figure}

\begin{figure}[!h]
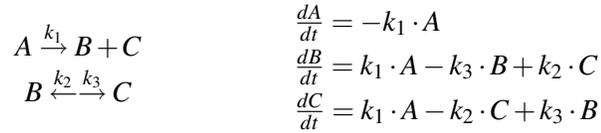

\begin{center}
\begin{minipage}{3cm}
$$\begin{array}{c}
A \ra{k_1} B + C\\
B \la{k_2}\ra{k_3} C
\end{array}
$$
\end{minipage}
\hspace{1cm}
\begin{minipage}{3cm}
$$\begin{array}{l}
\frac{dA}{dt} = -k_1\cdot A\\[1mm]
\frac{dB}{dt} = k_1\cdot A - k_3\cdot B + k_2\cdot C\\[1mm]
\frac{dC}{dt} = k_1\cdot A - k_2\cdot C + k_3\cdot B
\end{array}
$$
\end{minipage}
\end{center}
\caption{Example of an affine system}
\label{fig:affine}
\end{figure}

Multi-affine system is achieved from the system of biochemical
reactions by employing the law of mass action. In
Figure~\ref{fig:multiaffine} there is an example of a simple
biochemical system represented mathematically as a multi-affine
system. If all the reactions are of the first-order, in particular,
the number of reactants in each reaction is at most one, then the
system falls into a specific subclass of dynamic systems -- the
resulting ODE system is affine. An example of an affine system is
given in Figure~\ref{fig:affine}.

\subsection{Abstraction Procedure}

\begin{figure}
\begin{center}
\begin{minipage}{4cm}
$A\la{k_1}\ra{k_2} B$\\[3mm]
$\frac{dA}{dt}=-k_1\cdot A + k_2\cdot B$\\[1mm]
$\frac{dB}{dt}= k_1\cdot A - k_2\cdot B$\\[3mm]
thresholds on $A$: $0,5,6,10$\\
thresholds on $B$: $0,2,3,5$
\end{minipage}
\hspace{1cm}
\begin{minipage}{7cm}
\includegraphics[scale=.2]{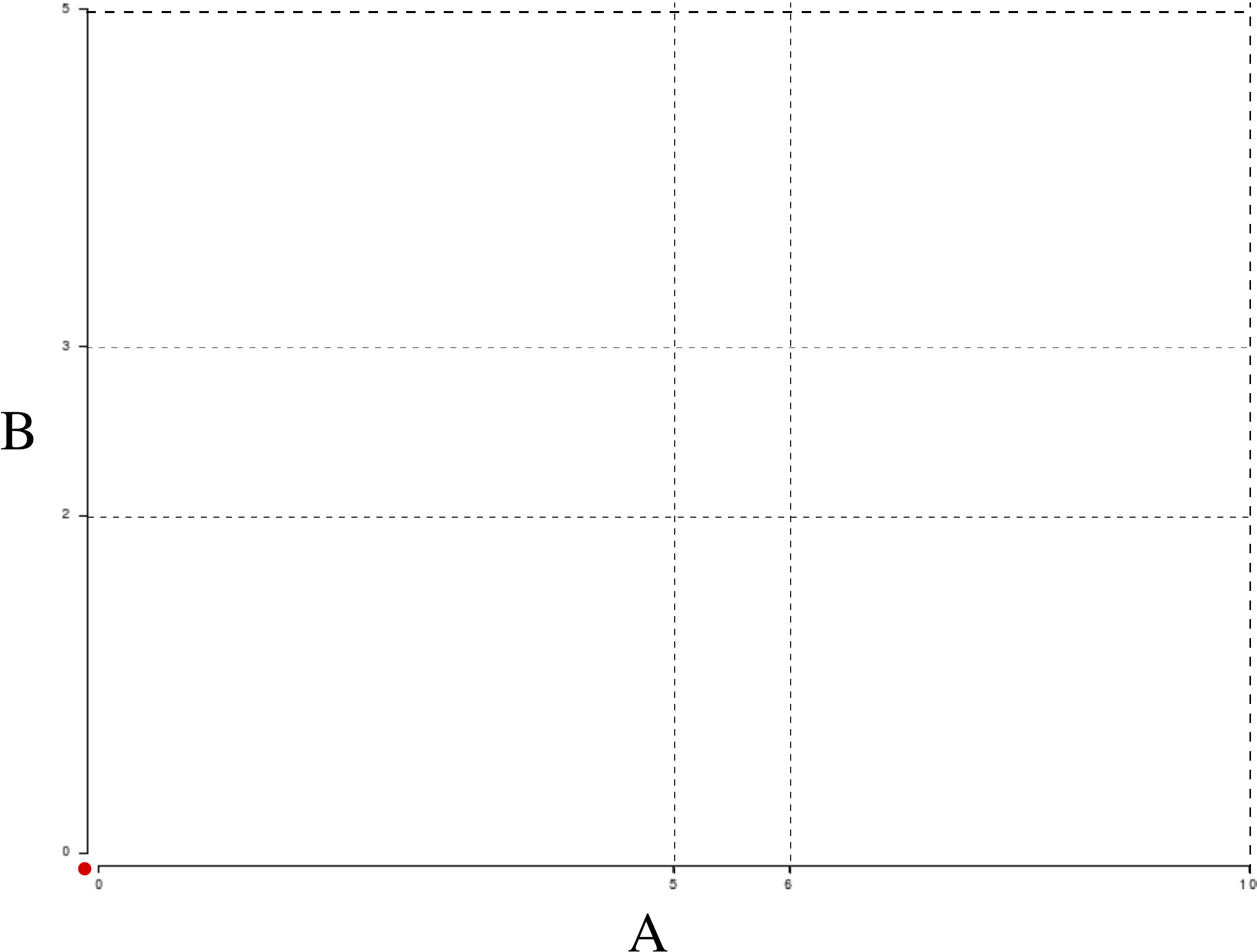}
\end{minipage}
\end{center}
\caption{Example of a rectangular partition of a two-dimensional system}
\label{fig:partition}
\end{figure}

The rectangular abstraction method employed in BioDiVinE relies on
results of de Jong, Gouz\'e~\cite{Hidde} and Belta, Habets~\cite{BH06}.
Each variable is assigned a set of specific (arbitrarily defined)
points, so-called \emph{thresholds}, expressing concentration levels
of special interest. This set contains two specific thresholds -- the
maximal and the minimal concentration level. Existence of these two
thresholds comes from the biophysical fact that in any living organism
each biochemical species cannot unboundedly increase its
concentration. The intermediate thresholds then define a partition of
the (bounded) continuous state space. The individual regions of the
partition are called \emph{rectangles}. An example of a partition is
given in Figure~\ref{fig:partition}.

\begin{figure}
\begin{center}
\includegraphics[scale=.4]{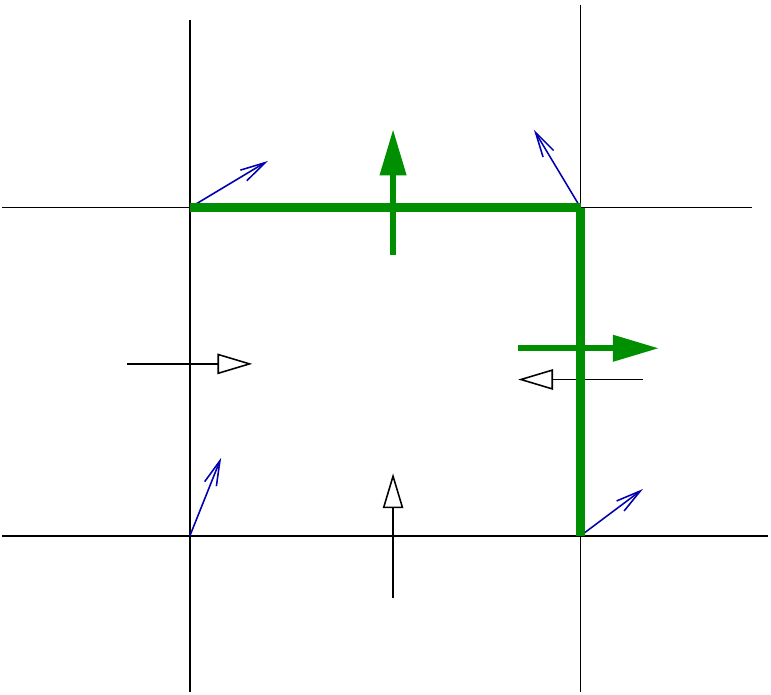}
\end{center}
\caption{Intuition behind the construction of the rectangular abstraction transition system}
\label{fig:moving}
\end{figure}

\begin{figure}
\begin{center}
\includegraphics[scale=.25]{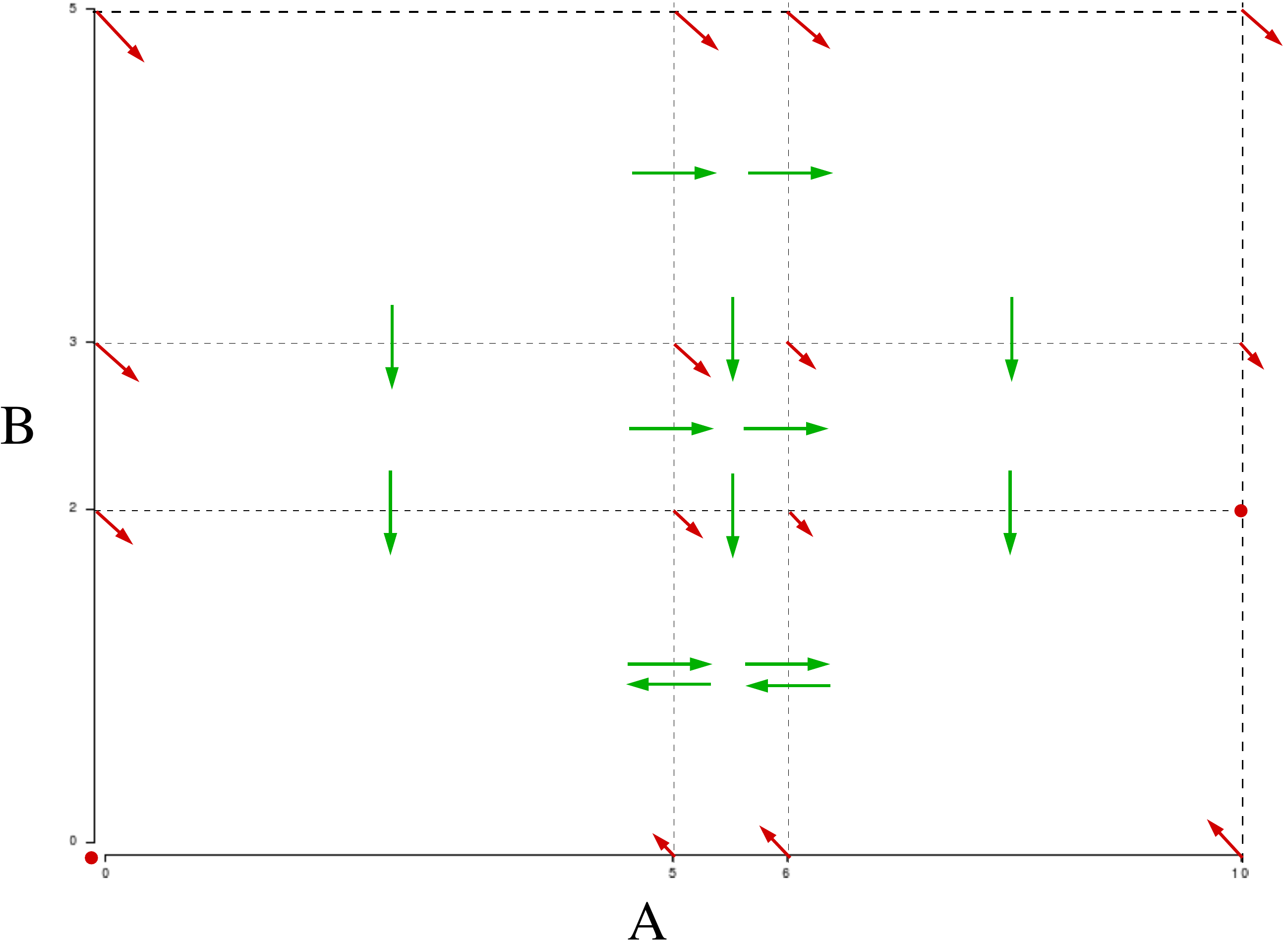}$\quad\quad$
\includegraphics[scale=.25]{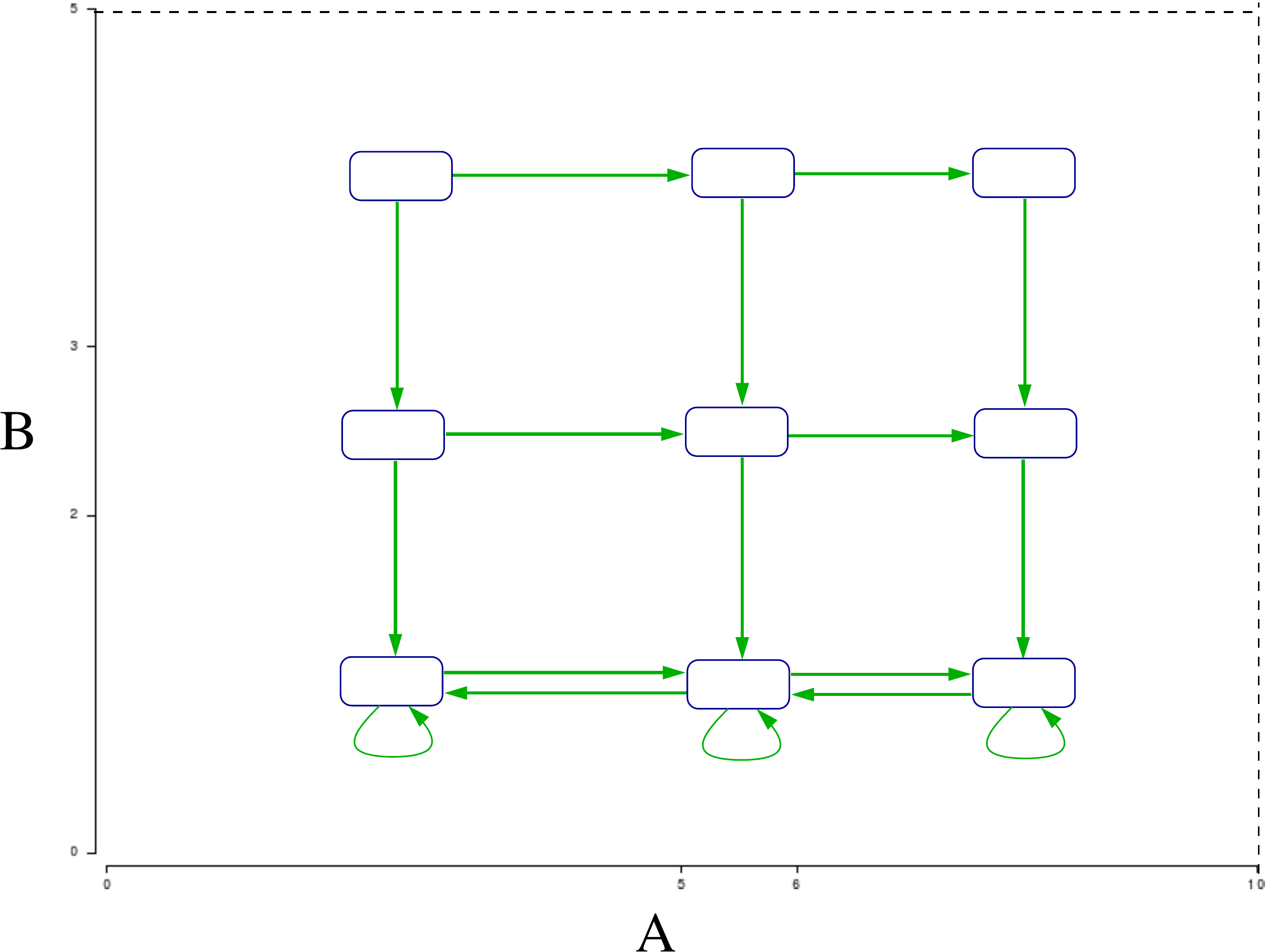}
\end{center}
\caption{Example of a rectangular abstraction transition system}
\label{fig:exampleaffine}
\end{figure}

The partition of the system gives us directly the finite discrete
abstraction of the dynamic system. In particular, BioDiVinE implements
a (discrete) state space generator that constructs a finite automaton
representing the rectangular abstraction of the system dynamics.
Since the states of the automaton are made directly by the rectangles,
the automaton is called \emph{rectangular abstraction transition
  system} (RATS). The algorithm for the state space generator of RATS
has been presented in~\cite{HIBI09}. The idea behind this algorithm is
based on the results~\cite{Hidde,BH06}. The main point is that for each rectangle the exit
faces are determined. The intuition is depicted in
Figure~\ref{fig:moving}. There is a transition from a rectangle to its
neighbouring rectangle only if in the vector field considered in the
shared face there is at least one vector whose particular component
agrees with the direction of the transition. The important result is
that in a multi-affine system it suffices to consider only the vector
field in the vertices of the face. In Figure~\ref{fig:moving}, the
exit faces of the central rectangle are emphasised by bold lines.  In
Figure~\ref{fig:exampleaffine} there is depicted the rectangular
abstraction transition system constructed for the affine system from
Figure~\ref{fig:partition}. It is known that the rectangular
abstraction is an overapproximation with respect to trajectories of
the original dynamic system.

There is one specific issue when considering the time progress of the
abstracted trajectories. If there exists a point in a rectangle from
which there is no trajectory diverging out through some exit face,
then there is a self-transition defined for the rectangle. In
particular, this situation signifies an equilibrium inside the
rectangle. Such a rectangle is called non-transient. For affine
systems there is known a sufficient and necessary condition that
characterises non-transient rectangles by the vector field in the
vertices of the rectangle.  However, for the case of multi-affine
system, only the necessary condition is known. Hence, for multi-affine
systems BioDiVinE can treat as non-transient some states which are not
necessarily non-transient.

\subsection{Model Checking}

In the field of formal verification of software and hardware systems,
model checking refers to the problem of automatically testing whether
a simplified model of a system (a finite state automaton) meets a
given specification.  Specification is stated by means of a temporal
logic formulae.  In the setting of RATS, model checking can be used in
two basic ways:
\begin{enumerate}
\item to automatically detect presence of particular dynamics
  phenomena in the system
\item to verify correctness of the model (i.e., checking whether some
  undesired property is exactly avoided)
\end{enumerate}

In the case of dynamic systems we use Linear Temporal Logic (LTL)
(see~\cite{MUB-MC} for details on LTL syntax and semantics interpreted
on automata). LTL can be also directly interpreted on trajectories of
dynamic systems (see e.g.~\cite{MUB-Clarke} for definition of the
semantics). Given a dynamic system $S$ with a particular initial state
we can then say that $S$ satisfies a formula $\varphi$, written
$S\models\varphi$, only if the trajectory starting at the initial
state satisfies $\varphi$. In the context of automata, LTL logic is
interpreted universally provided that a formula $\varphi$ is satisfied
by the automaton $A$, written $A\models \varphi$, only if each
execution of the automaton starting from any initial state satisfies
$\varphi$.  The following theorem characterises the relation between
validity of $\varphi$ in the rectangular abstraction automaton and in
the original dynamic system.

\begin{theorem}
  Consider a dynamic system $S$ and the associated RATS $A$. If $A\models\varphi$ then $S\models\varphi$.
\end{theorem}

The theorem states
that when model checking of a particular property on a RATS returns
true, we are sure that the property is satisfied in the original
dynamic system.  However, when the result is negative, the
counterexample returned does not necessarily reflect any trajectory in
the original system.

\begin{figure}
\begin{center}
\includegraphics[scale=.25]{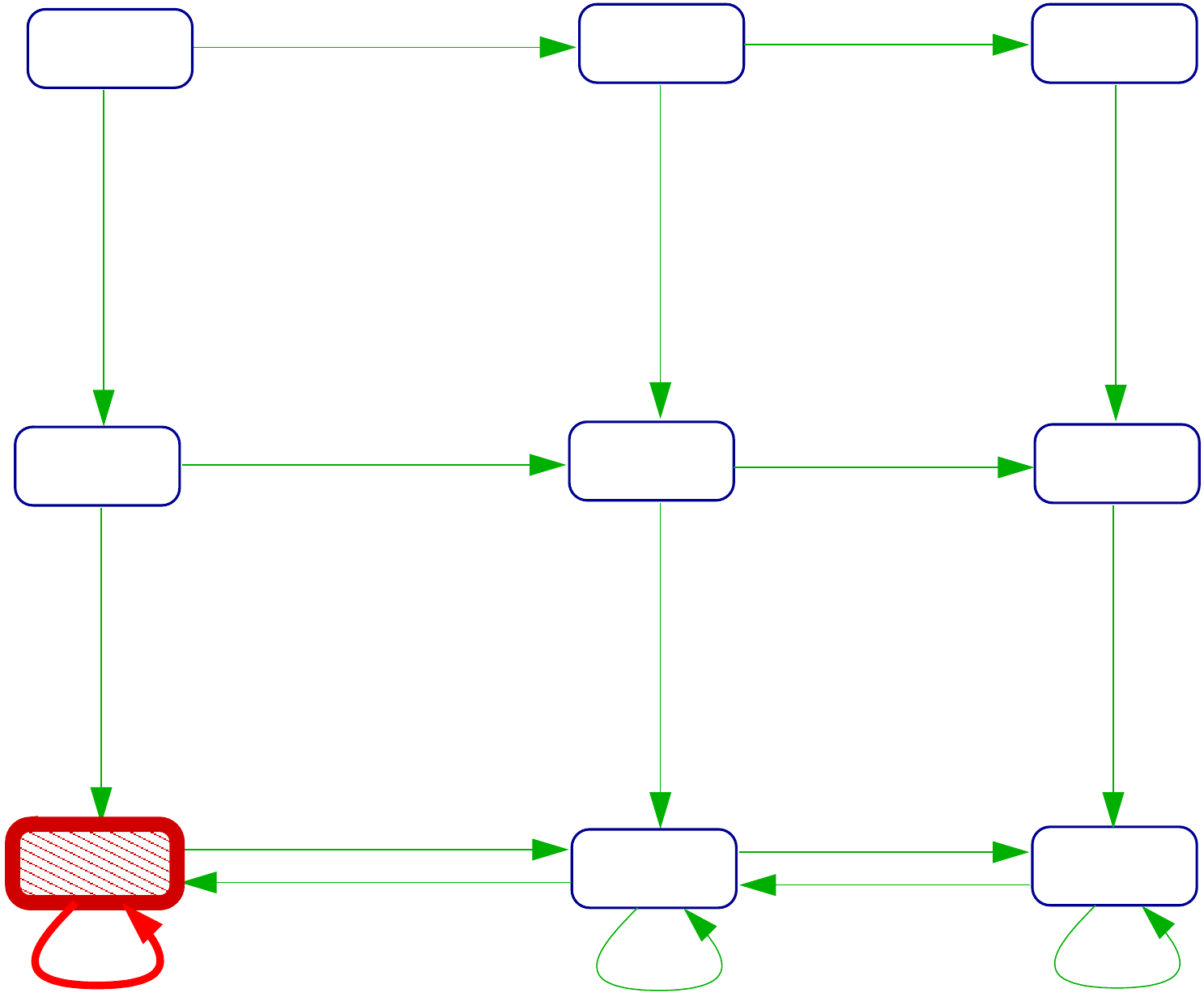}
\end{center}
\caption{A counterexample contradicting the property of reaching $B>3$ in finite time}
\label{fig:counterex}
\end{figure}

The system in Figure~\ref{fig:exampleaffine} satisfies a formula $\F\G
(B\leq 3)$ expressing the temporal property stating that despite the
choice of the initial state the system eventually stabilises at states
where concentration of $B$ is kept below $3$. Now let us consider a
formula $\F (B>3)$ expressing the property that despite the initial
settings, the concentration of $B$ will eventually exceed the
concentration level $3$. In this case the model checking returns one
of the counterexamples as emphasised in Figure~\ref{fig:counterex}
stating that if initially $A<5$ and $B<3$ then $B$ is not increased
while staying indefinitely long in the shaded state.




\section{BioDiVinE Tool}
\label{sec:biodivine}



BioDiVinE employs aggregate power of network-interconnected
workstations (nodes) to analyse large-scale state transition systems whose exploration is beyond capabilities of
sequential tools. System properties can be specified either directly in Linear
Temporal Logic (LTL) or alternatively as processes describing undesired
behaviour of systems under consideration (negative claim automata).
From the algorithmic point of view, the tool implements a
variety of novel parallel algorithms~\cite{DVEowcty,BBC05} for cycle detection (LTL model checking). By these algorithms, the entire state space is uniformly split into partitions and every partition is distributed to a particular computing node. Each node is responsible for generating the respective state-space partition on-the-fly while storing visited states into the local memory.

\begin{figure}
\begin{center}
\includegraphics[width=2.8in]{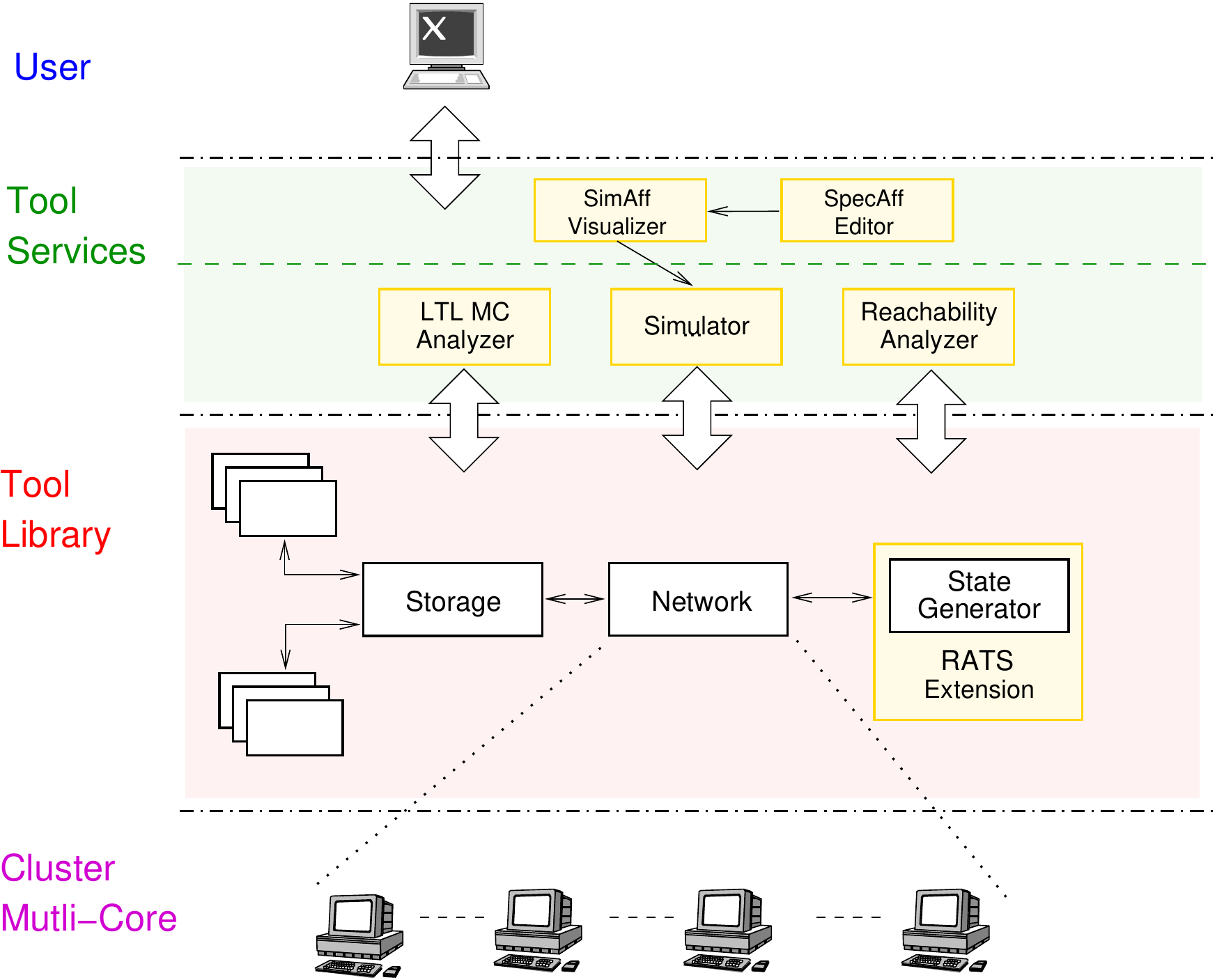}
\end{center}
\caption{BioDiVinE Toolset Architecture}
\label{fig:toolarch}
\end{figure}

The state space generator constructs the rectangular abstraction transition system for a given multi-affine system. The scheme of the tool architecture is provided in Figure~\ref{fig:toolarch}.
Library-level components are responsible for constructing, managing and distributing the state space. They form the core of the tool. The tool provides two graphical user interface components \emph{SpecAff} --- allowing editing of biological models in terms of chemical reactions, and \emph{SimAff} --- allowing visualisation of the simulation results.

\enlargethispage*{4mm}

\begin{figure}
\begin{center}
\includegraphics[width=2.24in,height=1.36in]{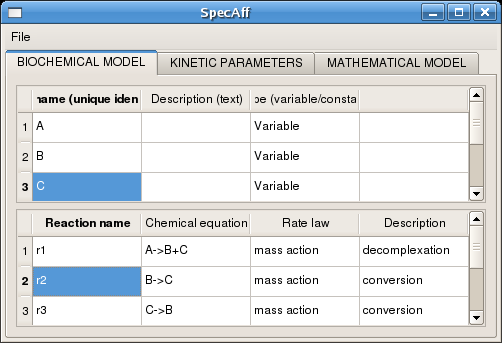}
\end{center}
\caption{A biochemical model specified in BioDiVinE GUI}
\label{fig:specaffgui}
\end{figure}

The input (biochemical) model is given by the following data:
\begin{itemize}
\item list of chemical species,
\item list of partitioning thresholds given for each species,
\item list of chemical reactions.
\end{itemize}

\begin{figure}
\begin{center}
\hspace*{1cm}\begin{minipage}{4cm}
{\scriptsize
\begin{verbatim}
VARS:A,B,C

EQ:dA = (-0.1)*A
EQ:dB = 0.1*A + (-1)*B + 1*C
EQ:dC = 0.1*A + 1*B + (-1)*C

TRES:A: 0, 4, 6, 10
TRES:B: 0, 2, 10
TRES:C: 0, 2, 4, 10

INIT:   4:6, 0:2, 2:4
\end{verbatim}
}
\end{minipage}
\raisebox{-22mm}{\includegraphics[width=8cm]{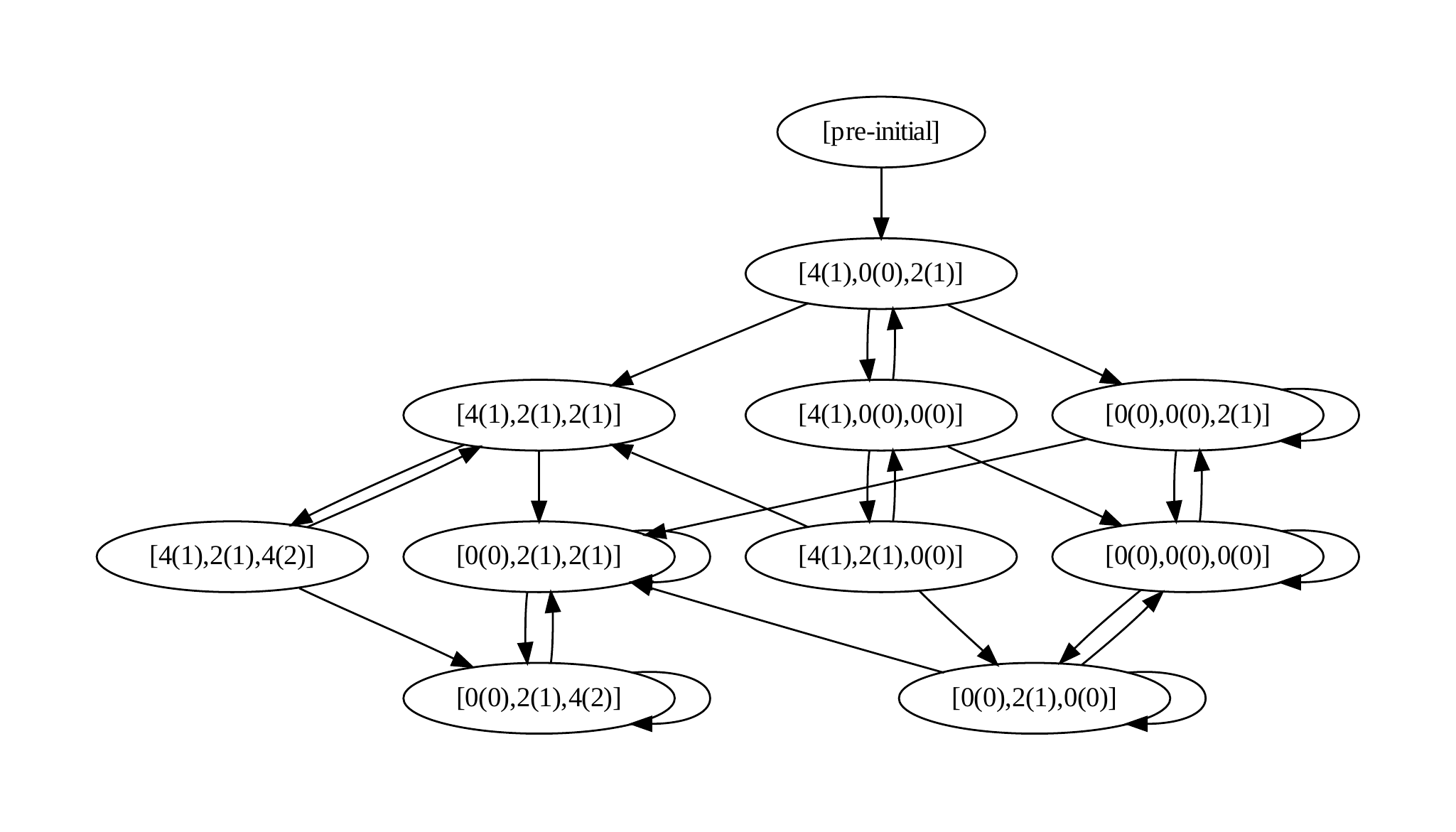}}
\end{center}
\caption{A multi-affine ODE model and its state space generated by BioDiVinE}
\label{fig:demobioexample}
\end{figure}

The biochemical model together with the tested property and initial conditions are then automatically translated into a multi-affine system of ODEs forming the mathematical model that can be analysed by BioDiVinE algorithms. The mathematical model consists of the following data:

\begin{itemize}
\item list of variables,
\item list of (multi-affine) ODEs,
\item list of partitioning thresholds given for each species,
\item list of initial rectangular subspaces (the union of these subspaces forms the initial condition),
\item B\"uchi automaton representing an LTL property (this data is not needed for simulation).
\end{itemize}

\begin{figure}
\begin{center}
\includegraphics[width=4.5cm]{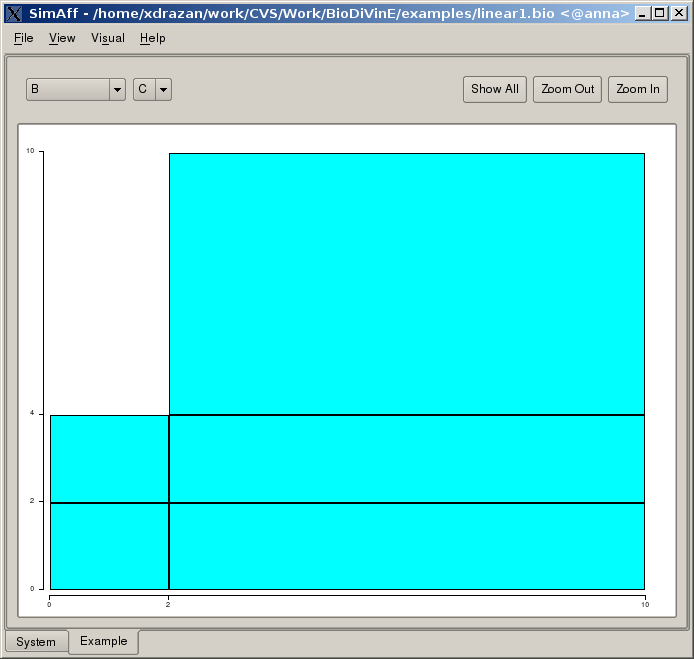}
\end{center}
\caption{Visualisation of the state space in the BioDiVinE GUI, states are projected onto the BC concentration plane}
\label{fig:simaffgui}
\end{figure}

An example of a simple three-species model representing three biochemical reactions $A \rightarrow B + C$, $B \rightleftharpoons C$ is showed in Figure~\ref{fig:specaffgui}.
The first reaction is performed at rate $0.1~s^{-1}$. The second two reversible reactions are at rate $1~s^{-1}$.
The respective mathematical model is showed in Figure~\ref{fig:demobioexample} on the left in the textual \texttt{.bio} format.
For each variable there is specified the equation as well as the list of real values representing individual threshold positions. The initial condition is defined in this particular case by a single rectangular subspace: $A\in\langle 4,6\rangle, B\in\langle 0,2\rangle$, and $C\in\langle 2,4\rangle$. The state space generated for this setting is depicted in Figure~\ref{fig:demobioexample} on the right.
Figure~\ref{fig:simaffgui} demonstrates visualisation features of the BioDiVinE GUI.


\begin{figure}
\begin{center}
\includegraphics[width=.48\textwidth]{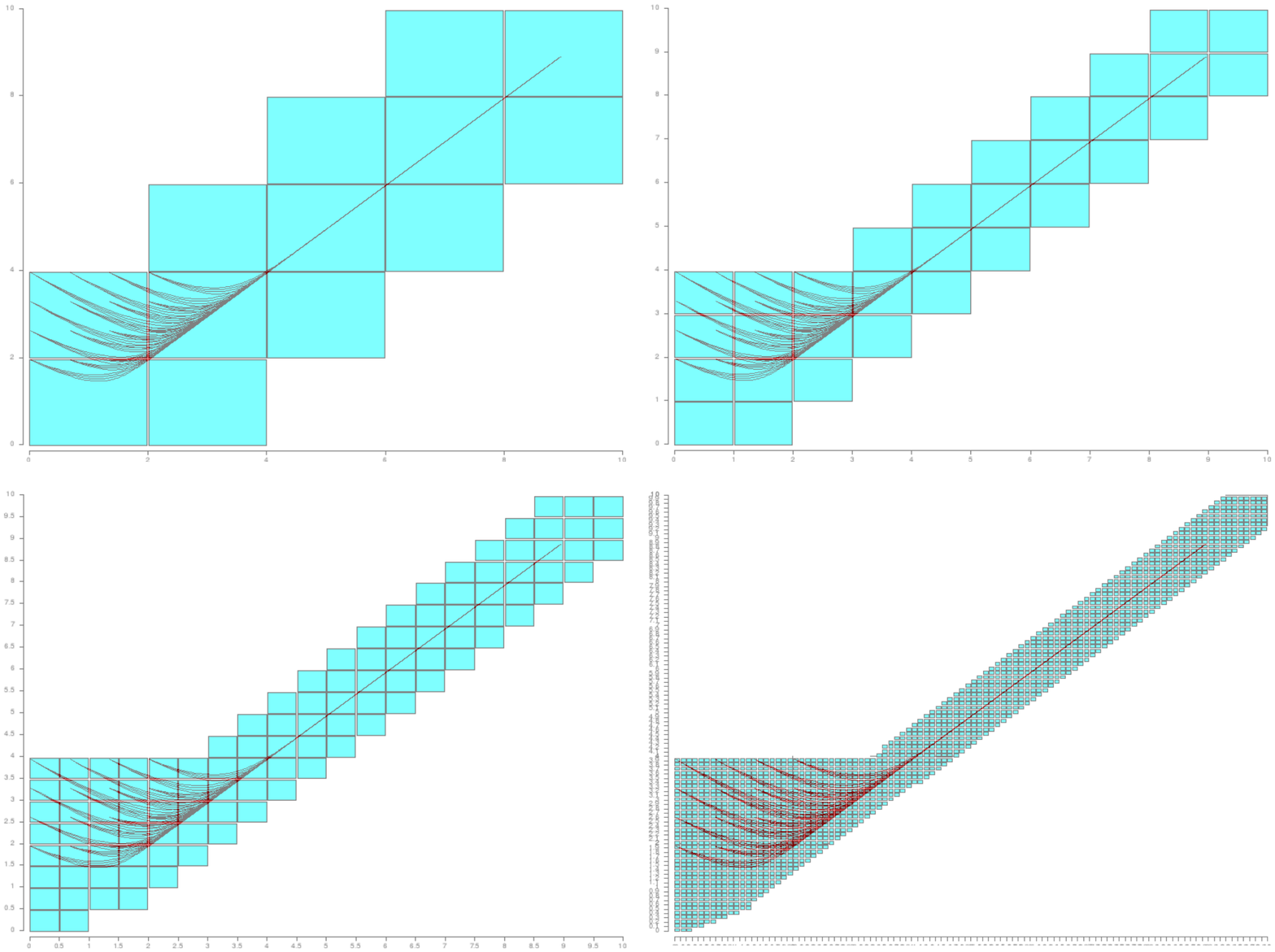}
\includegraphics[width=.48\textwidth]{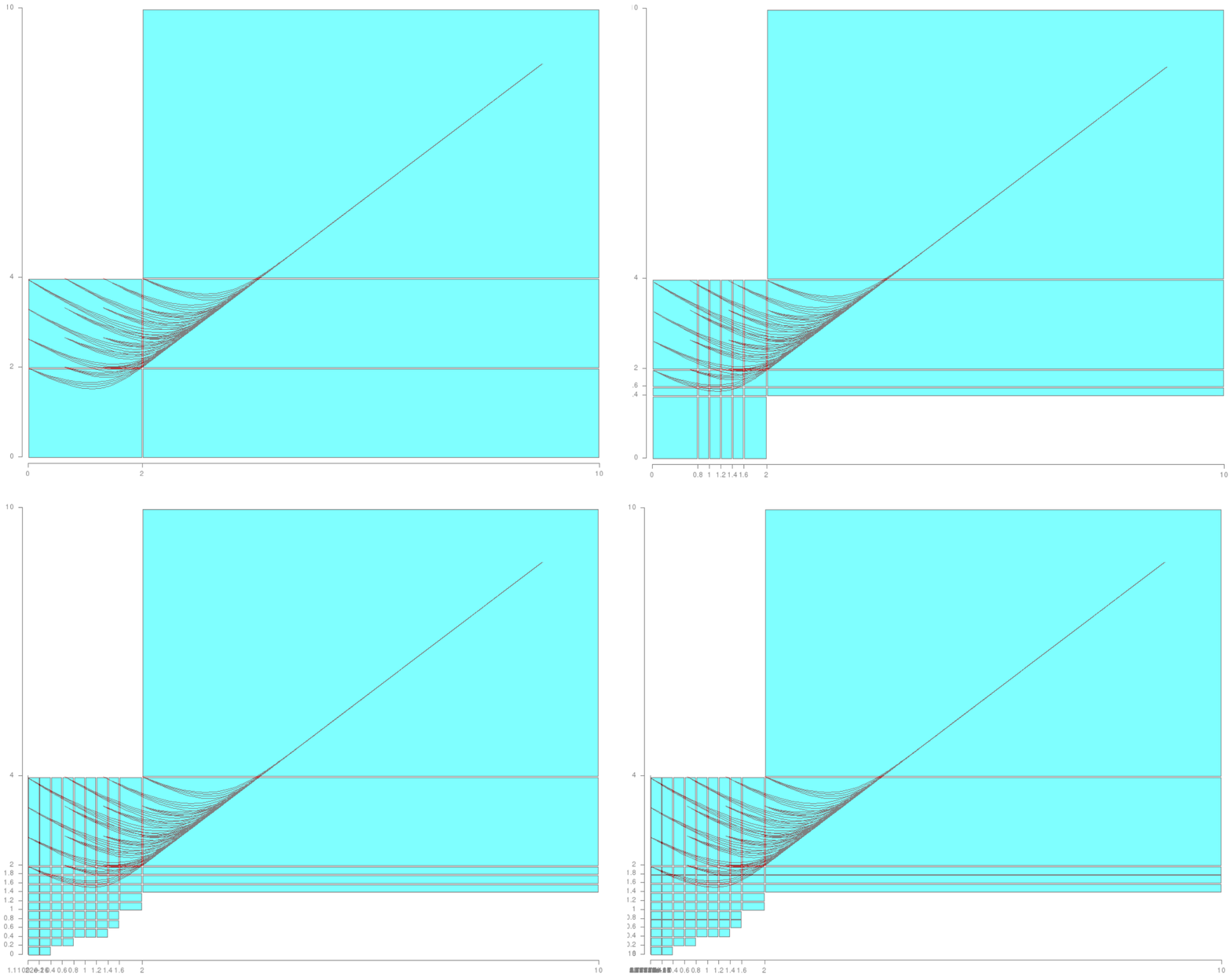}
\end{center}
\caption{A visualisation of the state-spaces in BC projection with uniform (left) and automatic (right) gradual threshold refinement, manually augmented by numeric simulation trace from COPASI~\cite{COPASI}}
\label{fig:refinement}
\end{figure}

\begin{figure}
\begin{center}
{\scriptsize
\begin{minipage}{4.0cm}
\begin{verbatim}
=======================
--- Accepting cycle ---
=======================
[pre-initial]
[4(1),0(0),2(1)-PP:0]
[4(1),2(1),2(1)-PP:0]
[4(1),2(1),3(2)-PP:0]
[4(1),4(2),3(2)-PP:1]
[0(0),4(2),3(2)-PP:1]
======= Cycle =======
[0(0),2(1),3(2)-PP:1]
[0(0),2(1),2(1)-PP:1]


\end{verbatim}
\end{minipage}
\begin{minipage}{4.8cm}
\begin{verbatim}
==============================
  --- No accepting cycle ---
==============================
states:            33
transitions:       81
iterations:        1
size of a state:   16
size of appendix:  12
cross transitions: 0
all memory         56.5 MB
time:              0.115177 s
-------------------
0: local states:      33
0: local memory:      56.5
\end{verbatim}
\end{minipage}
\begin{minipage}{6.5cm}
\includegraphics[width=6.5cm]{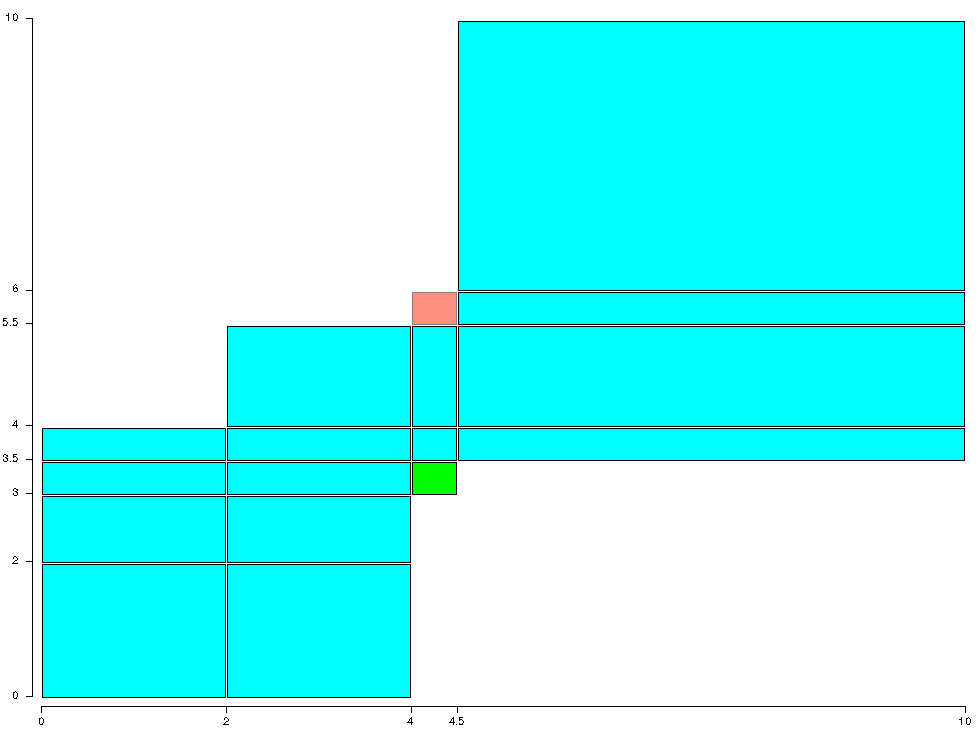}
\end{minipage}
}
\end{center}
\caption{Results of model-checking for property 1 (left), property 2 (middle), visualisation of reachable (green) and unreachable (red) regions}
\label{fig:modelchecking}
\end{figure}

For model checking analysis, BioDiVinE relies on the parallel LTL model checking algorithms of the underlying DiVinE library~\cite{DVEtool}. A given LTL formula is translated into a B\"uchi automaton which represents its negation. That way the automaton represents the never claim property. The automaton is automatically generated for an LTL formula and merged with the mathematical model by the \texttt{divine.combine} utility. An example of a model extended with a never claim property is showed in Figure~\ref{fig:property}. In particular, the first automaton \texttt{LTL\_property1} specified in terms of the DiVinE language represents a never claim for the safety LTL formula
$\mathbf{G} \neg (B\geq4 \wedge B\leq4.5 \wedge C\geq3 \wedge C\leq3.5)$
expressing that concentrations of species $B$ and $C$ will never enter the specified rectangular region. The unreachability of a slightly different region is defined by the automaton for property~2. The results of the model-checking procedure are showed in Figure~\ref{fig:modelchecking}. Property~1 has been proved false by finding a run of the system visiting the specified region (states of run given as list), while property~2 has been proved true by extensively searching all the system runs and not finding any one that would cross the region specified in property~2.

\begin{figure}
\begin{center}
{\scriptsize
\begin{minipage}{4.8cm}
\begin{verbatim}
VARS:A,B,C

EQ:dA = (-0.1)*A
EQ:dB = 0.1*A + (-1)*B + 1*C
EQ:dC = 0.1*A + (1)*B + (-1)*C

TRES:A: 0, 4, 6, 10
TRES:B: 0, 2, 10
TRES:C: 0, 2, 4, 10

INIT:   4:6, 0:2, 2:4
\end{verbatim}
\end{minipage}
\begin{minipage}{5.7cm}
\begin{verbatim}
process LTL_property1 {
state q1, q2;
init q1;
accept q2;
trans
q1 -> q2 { guard B>4 && B<4.5
              && C>3 && C<3.5; },
q1 -> q1 {},
q2 -> q2 {};
}

system sync property LTL_property1;
\end{verbatim}
\end{minipage}
\begin{minipage}{5cm}
\begin{verbatim}
process LTL_property2 {
state q1, q2;
init q1;
accept q2;
trans
q1 -> q2 { guard B>4 && B<4.5
              && C>5.5 && C<6; },
q1 -> q1 {},
q2 -> q2 {};
}

system sync property LTL_property2;
\end{verbatim}
\end{minipage}
}
\end{center}
\caption{A multi-affine model extended with a never claim automaton for property 1 and property 2}
\label{fig:property}
\end{figure}

The choice of threshold values for each variable affects greatly the shape and size of the generated state-space. The refinement of a given partitioning --- the addition of more thresholds to a set of former thresholds --- may result in the unreachability of a part of a region reachable before.


Since manual refinement of thresholds by adding numeric values can be tedious or unintuitive, two more advanced methods are available in BioDiVinE. The first method divides a given interval uniformly into subintervals of a given size, while the second method tries a more sophisticated automatic technique of dividing regions according to signs of the concentration derivatives inside these regions~\cite{KB08}.
The state-spaces resulting from the gradual application of both threshold refinement approaches are depicted in Figure~\ref{fig:refinement}. It is important to mention that the overall size of the state-space depends exponentially on the number of thresholds for all species. However, in some cases the actual reachable subspace of the whole state-space may be only polynomial in the number of thresholds.

For any multi-affine model extended with a never claim automaton as showed in Figure~\ref{fig:property}, the parallel model checking algorithms can be directly called.

\begin{figure}
\begin{center}
$$S + E\rightleftharpoons ES_1\rightleftharpoons ES_2\rightleftharpoons\cdots\rightleftharpoons ES_k\rightarrow P + E$$

\parbox{6cm}{
$$\begin{array}{rcl}
\dot{S}&=& ES_1 -0.01\cdot E\cdot S\\
\dot{E}&=& ES_1 - 0.01\cdot S\cdot E + ES_1\\
\dot{ES_1}&=& 0.01\cdot E\cdot S - ES_1 - ES_1\\
\dot{ES_2}&=& \cdot ES_1 - 2\cdot ES_2 + ES_3\\
&.&\\
&.&\\
&.&\\
\dot{ES}_{k-1} &=& ES_{k - 2} - 2\cdot ES_{k-1}  + ES_k\\
\dot{ES}_k &=& ES_{k-1} - 2\cdot ES_k
\end{array}$$
}
\raisebox{-3cm}{
\includegraphics[width=8.0cm]{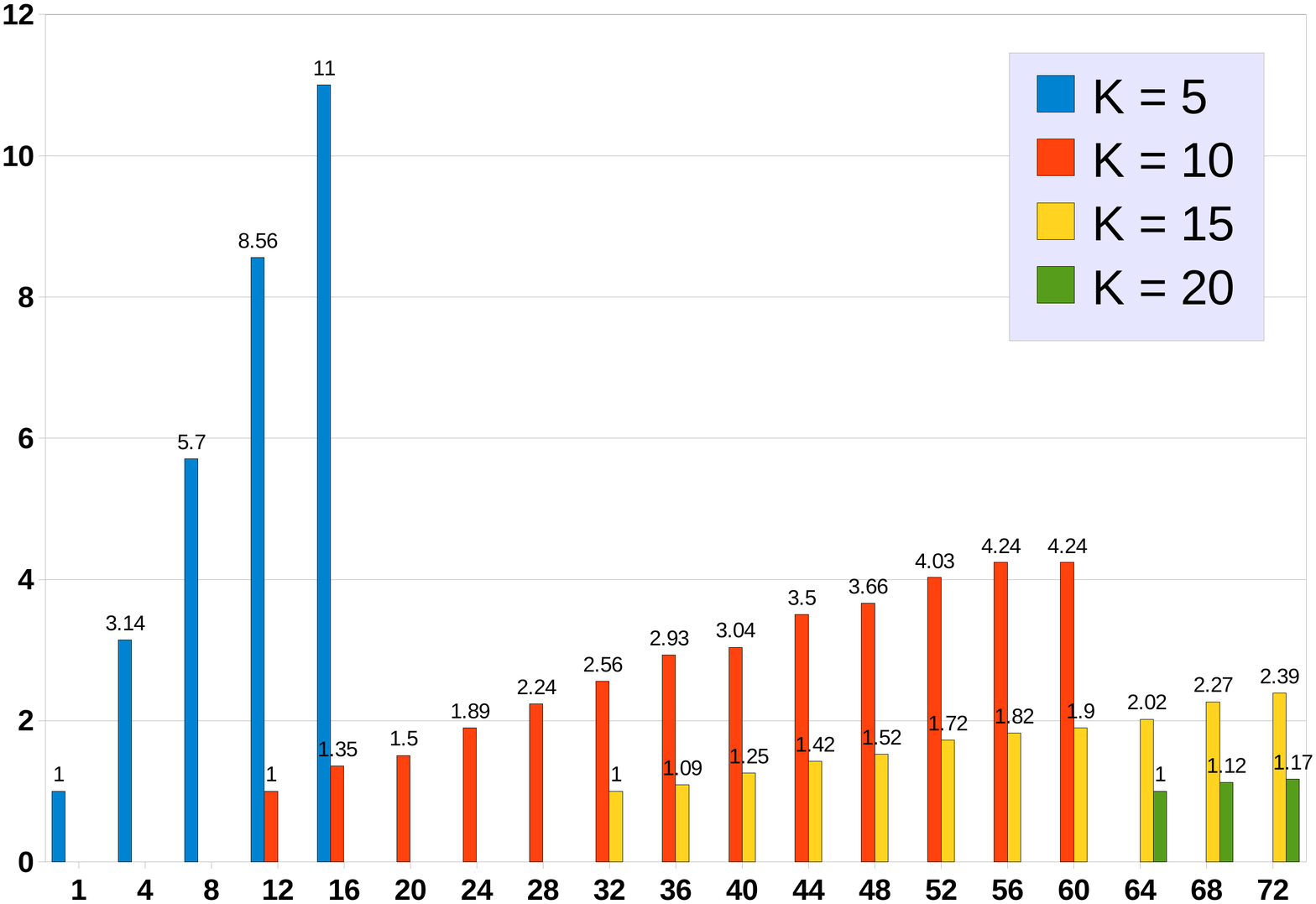}}
\end{center}
\caption{Scaling of model checking algorithms on a homogeneous cluster, X-axis shows number of cluster nodes, Y-axis gives speedup relative to system size and number of nodes}
\label{fig:experiments}
\end{figure}

We have performed several experiments~\cite{HIBI09} in order to show scaling of the algorithms when distributed on several cluster nodes. Figures~\ref{fig:mmexps} and \ref{fig:experiments} show scaling of model checking conducted on a simple model of a reaction network representing a catalytic reaction scaled for different numbers of intermediate products.

\begin{figure}
\begin{center}
\scalebox{0.80}{
{\scriptsize
\begin{tabular}{|c||c|c||ccccccccccccccccccc|}
\hline
    &                 &                 & \multicolumn{19}{c|}{Time on particular number of cluster processor cores (s)} \\\cline{4-22}
$k$ & States          & Trans           & 1      & 4      & 8      & 12     & 16     & 20     & 24     & 28     & 32     & 36     & 40     & 44     & 48     & 52     & 56     & 60     & 64  & 68 & 72 \\\hline
5   & $3\cdot 10^4$   & $8.5\cdot 10^4$ & 15.4   & 4.9    & 2.7    & 1.8    & 1.4    & $\bot$ & $\bot$ & $\bot$ & $\bot$ & $\bot$ & $\bot$ & $\bot$ & $\bot$ & $\bot$ & $\bot$ & $\bot$ & $\bot$&  $\bot$& $\bot$ \\
    &  relative       & speedup         & 1      & 3.14   & 5.7    & 8.56   & 11     & $\bot$ & $\bot$ & $\bot$ & $\bot$ & $\bot$ & $\bot$ & $\bot$ & $\bot$ & $\bot$ & $\bot$ & $\bot$ & $\bot$&  $\bot$& $\bot$ \\
10  & $9\cdot 10^5$   & $3.2\cdot 10^6$ & $\top$ & $\top$ & $\top$ & 161    & 119    & 107    & 85     & 72     & 63     & 55     & 53     & 46     & 44     & 40     & 38     & 38    & $\bot$ & $\bot$ & $\bot$ \\
    &  relative       & speedup         & $\top$ & $\top$ & $\top$ & 1      & 1.35   & 1.5    & 1.89   & 2.24   & 2.56   & 2.93   & 3.04   & 3.5    & 3.66   & 4.03   & 4.24   & 4.24  & $\bot$ & $\bot$ & $\bot$ \\
15  & $1.6\cdot 10^6$ & $6.5\cdot 10^6$ & $\top$ & $\top$ & $\top$ & $\top$ & $\top$ & $\top$ & $\top$ & $\top$ & 222    & 204    & 177    & 156    & 146    & 129    & 122    & 117    & 110 & 98 & 93 \\
    &  relative       & speedup         & $\top$ & $\top$ & $\top$ & $\top$ & $\top$ & $\top$ & $\top$ & $\top$ & 1      & 1.09   & 1.25   & 1.42   & 1.52   & 1.72   & 1.82   & 1.9    & 2.02 & 2.27 & 2.39 \\
20  & $3.2\cdot 10^6$ & $1.4\cdot 10^7$ & $\top$ & $\top$ & $\top$ & $\top$ & $\top$ & $\top$ & $\top$ & $\top$ & $\top$ & $\top$ & $\top$ & $\top$ & $\top$ & $\top$ & $\top$ & $\top$ & 202 & 180 & 173 \\
    &  relative       & speedup         & $\top$ & $\top$ & $\top$ & $\top$ & $\top$ & $\top$ & $\top$ & $\top$ & $\top$ & $\top$ & $\top$ & $\top$ & $\top$ & $\top$ & $\top$ & $\top$ & 1   & 1.12 & 1.17 \\\hline
\end{tabular}
}}
\end{center}
\caption{Scaling of model checking algorithms on a homogeneous cluster}
\label{fig:mmexps}
\end{figure}

\enlargethispage*{5mm}

\section{Case Study: Ammonium Transport in E. Coli}
\label{sec:casestudy}



In this section we present a case study conducted using the current version of BioDiVinE. Since the rectangular abstraction method of multi-affine systems implemented in BioDiVinE is a relatively new result of applied control theory, its application is still in the stage of experimentation. In fact, we are still unaware of any case studies that apply this method to real biological models. In this case study we focus on demonstrating the usability of rectangular abstraction to analysis of biological models. To this end, we consider a simple biological model that specifies ammonium transport from external environment into
the cells of Escherichia Coli. This simplified model is based on a published model of the E. Coli ammonium assimilation system~\cite{ECMOANMODEL}
which was developed under the EC-MOAN project (\texttt{http://www.ec-moan.org}).
The metabolic reactions and regulatory reactions in the original model were removed.

\subsection{Model Description}

\begin{figure}
\begin{center}
\includegraphics[scale=.5]{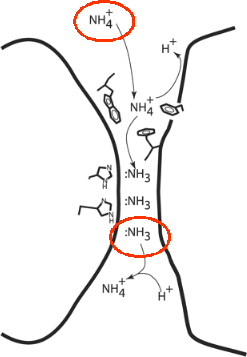}
\hspace{2cm}
\includegraphics[scale=.5]{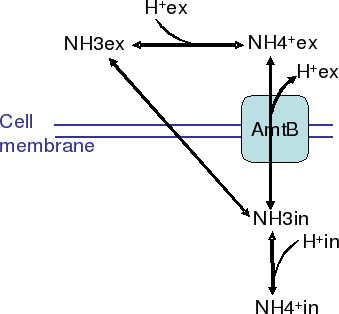}
\end{center}
\caption{E. Coli ammonium transport mechanism and the respective pathway}
\label{fig:pathway}
\end{figure}

E. coli can express membrane bound transport proteins for the transportion of small molecules from the environment into the cytoplasm at certain conditions. At normal ammonium concentration, the free diffusion of ammonia can provide enough flux for the growth requirement of nitrogen. When ammonium concentration is very low, E. coli cells express $AmtB$ (an \emph{ammonia transporter}) to complement the deficient diffusion process. Three molecules of $AmtB$ (trimer) form a channel for the transportation of ammonium/ammonia. Protein structure analysis revealed that $AmtB$ binds $NH4^+$ at the entrance gate of the channel, deprotonates it and conducts $NH_3$ into the cytoplasm as illustrated in Figure~\ref{fig:pathway} (left)~\cite{science}. At the periplasmic side of the channel there is a wider vestibule site capable of recruiting $NH_4^+$ cations. The recruited cations are passed through the hydrophobic channel where the pKa of $NH_4^+$ was shifted from $9.25$ to below $6$, thereby shifting the equilibrium toward the production of $NH_3$. $NH_3$ is finally released at the cytoplasmic gate and converted to $NH_4^+$ because intracellular pH ($7.5$) is far below the pKa of $NH_4^+$.

In addition to the above mentioned $AmtB$ mediated transport,
the bidirectional free diffusion of the uncharged ammonia through the membrane is also included in the simplified model. The intracellular $NH_4^+$ is then metabolised by Glutamine Synthetase (GS). The whole model is depicted in Figure~\ref{fig:pathway} (right). The external ammonium is
represented in the uncharged and charged forms denoted $NH_3ex$ and $NH_4^+ex$. Analogously, the internal
ammonium forms are denoted $NH_3in$ and $NH_4^+in$. The biochemical model that combines $AmtB$ transport
with $NH_3$ diffusion is given in Table~\ref{tab:biochem}.
The kinetic parameters are based on the values in the original model.

\begin{table}
$$\begin{array}{c@{\hspace{1cm}}l}
AmtB + NH_4ex \la{k_1}\ra{k_2} AmtB:NH_4 & k_1=5\cdot 10^8, k_2=5\cdot 10^3\\
AmtB:NH_4 \ra{k_3} AmtB:NH_3 + H_{ex} & k_3 = 50\\
AmtB:NH_3 \ra{k_4} AmtB + NH_3in & k_4 = 50\\
NH_4in \ra{k_5} & k_5 = 80\\
NH_3in + H_{in} \la{k_6}\ra{k_7} NH_4in & k_6=1\cdot 10^{15}, k_7=5.62\cdot 10^{5}\\
NH_3ex \la{k_8}\ra{k_9} NH_3in & k_8=k_9=1.4\cdot 10^{4}
\end{array}
$$
\caption{Biochemical model of ammonium transport}
\label{tab:biochem}
\end{table}


By employing the law of mass action kinetics the reaction network is
transformed into the set of multi-affine ODEs as listed in
Table~\ref{tab:ammoniumODEs}. Since we are especially interested in
how the concentrations of internal ammonium change with respect to
the external ammonium concentrations, we employ the following simplifications:
\begin{itemize}
\item We do not consider the dynamics of the external ammonium forms, thus we take $NH_3ex$ and $NH_4^+ex$ as constants (the input parameters for the analysis).
\item We assume constant intracellular pH ($7.5$) and extracellular pH ($7.0$), thus
$H_{ex}$ and $H_{in}$ are calculated to be $3\cdot 10^{-8}$ and $10^{-7}$. Based on the extracellular pH and the total ammonium/ammonia concentration, concentrations of $NH_3ex$ and $NH_4^+ex$ can be calculated.
\end{itemize}
Without the loss of correctness, we simplify the notation of the cation $NH_4^+$ as $NH_4$.

\begin{table}
$$\begin{array}{c}
\frac{dAmtB}{dt} = -k_1\cdot AmtB\cdot NH_4ex + k_2\cdot AmtB:NH_4 + k_4\cdot AmtB:NH_3\\
\frac{dAmtB:NH_3}{dt} = k_3\cdot AmtB:NH_4 - k_4\cdot AmtB:NH_3\\
\frac{dAmtB:NH_4}{dt} = k_1\cdot AmtB\cdot NH_4ex - k_2\cdot AmtB:NH_4 - k_3\cdot AmtB:NH_4\\
\frac{dNH_3in}{dt} = k_4\cdot AmtB:NH_3 - k_7\cdot NH_3in + k_6\cdot NH_4in\\
\frac{dNH_4in}{dt} = k_5\cdot NH_4in + k_7\cdot NH_3in\cdot H_{in} - k_6\cdot NH_4in\\
\end{array}
$$
\caption{Mathematical model of ammonium transport}
\label{tab:ammoniumODEs}
\end{table}

\subsection{Model Analysis}

From the essence of biophysical laws, it is clear that the maximal
reachable concentration level accumulated in the internal ammonium
forms directly depends on the ammonium sources available in the
environment. However, it is not directly clear what particular maximal
level of internal ammonium is achievable at given amount of external
ammonium (distributed into the two forms). In the analysis we have
focused on just this phenomenon.
More precisely, the problem to solve has been to analyse how the setting of
the model parameters $NH_3ex$ and $NH_4^+ex$ affects the maximal
concentration level of $NH_3in$ and $NH_4^+in$ reachable from given
initial conditions.

During discussions with biologists we have found out that there is
currently not available any in vitro measurement of the $AmtB$-concentration (and also the concentrations of dimers $AmtB:NH_3$ and $AmtB:NH_4$). Hence there has
appeared the need to analyse the model with uncertain initial
conditions. Such a setting fits the current features of BioDiVinE,
especially the rectangular abstraction that naturally abstracts away the
exact concentration values up to the intervals between certain
concentration levels.

\subsubsection{Maximal reachable levels of internal ammonium forms}
\label{sec:maximalreach}

At first, we have considered the (abstracted) initial condition to be
set to the following intervals between concentration values:
$$
AmtB\in\langle 0,1\cdot 10^{-5}\rangle,\quad AmtB:NH_3\in\langle 0, 1\cdot 10^{-5}\rangle,\quad AmtB:NH_4\in\langle 0, 1\cdot 10^{-5}\rangle,$$\\[-12.5mm]
$$NH_3in\in\langle 1\cdot 1^{-6}, 1.1\cdot 10^{-6}\rangle,\quad
NH_4in\in\langle 2\cdot 10^{-6}, 2.1\cdot 10^{-6}\rangle$$

Note that the upper bounds as well as the initial intervals of
internal ammonium forms have been set with respect to the available
data obtained from the literature.

We have considered two rectangular partitions. The partition $(1)$ has
been set basically according to the initial conditions. The partition
$(2)$ has been obtained by running one iteration of the automatic
threshold refinement procedure to partition $(1)$. Numbers of
thresholds per each variable are given in Table~\ref{tab:partition1}.

\begin{table}
{\small
\begin{center}
\begin{tabular}{|c||ccccc|}
\hline
Partition & $AmtB$ & $AmtB:NH_3$ & $AmtB:NH_4$ & $NH_3in$ & $NH_4in$\\\hline
$(1)$ & $7$ & $3$ & $3$ & $6$ & $4$\\
$(2)$ & $7$ & $9$ & $3$ & $8$ & $26$\\\hline
\end{tabular}
\end{center}
}
\caption{Numbers of thresholds in partitions $(1)$ and $(2)$}
\label{tab:partition1}
\end{table}

We have conducted several model checking experiments in order to
determine the maximal reachable concentration levels of $NH_3in$ and
$NH_4^+in$. In particular, we have searched for the lowest $\alpha$
satisfying the property $\G (NH_3in<\alpha)$ and the lowest $\beta$
satisfying $\G (NH_4in<\beta)$. The property $\G p$ requires that all
paths available in the rectangular abstraction from the states
specified by the initial condition must satisfy the given proposition
$p$ at every state. Note that if the model checking method finds the
property $\G p$ false in the model, it also returns a counterexample
for that.  The counterexample satisfies the negation of the checked
formula, which is in this case $\F \neg p$. Interpreting this
observation intuitively for the given formulae, we use model
checking to find a path on which the species $NH_3in$ (resp. $NH_4in$)
exceeds the level $\alpha$ (resp. $\beta$).

We did not want to get precise values of $\alpha,\beta$, but we rather
wanted to get their good approximation. At the starting point, we
substituted for $\alpha$ (resp. $\beta$) the upper initial bounds of
the respective variables. Then we found the requested values by
iteratively increasing and decreasing $\alpha$ (resp. $\beta$). The
obtained results are summarised in Table~\ref{tab:results1}.

\begin{table}
{\small
\begin{center}
\begin{tabular}{|c|c|c|c||c|c|c|}
\hline
\multicolumn{4}{|c|}{Partition $(1)$} & \multicolumn{3}{|c|}{Partition $(2)$}\\
\hline
$\alpha$ & $\G (NH_3in<\alpha)$ & $\#$ states & Time ($\#$ nodes) & $\G (NH_3in<\alpha)$ & $\#$ states &  Time ($\#$ nodes)\\\hline
$1.1\cdot 10^{-6}$ & true & 1081 & 0.36 s (1) & true & $1.5\cdot 10^5$ & 1.9 s (18)\\
\hline\hline
$\beta$ & $\G (NH_4in<\beta)$ & $\#$ states & Time ($\#$ nodes) & $\G (NH_4in<\beta)$ & $\#$ states &  Time ($\#$ nodes)\\\hline
$1\cdot 10^{-3}$ & true & 2161 & 0.45 s (1) & true & $1.6\cdot 10^5$ & 2 s (18)\\
$5\cdot 10^{-4}$ & false & 4753 & 1.9 s (1) & false & $2.7\cdot 10^5$ & 3 s (18)\\
$6\cdot 10^{-4}$ & true & 2161 & 0.43 s (1) & true & $1.5\cdot 10^5$ & 1.8 s (18)\\
$5.4\cdot 10^{-4}$ & true & 1441 & 0.27 s (1) & true & $2.1\cdot 10^5$ & 4.2 s (18)\\
$5.3\cdot 10^{-4}$ & false & 3421 & 1.2 s (1) & false & $2.7\cdot 10^5$ & 2.2 s (18)\\
\hline
\end{tabular}
\end{center}
}
\caption{Experiments on detecting maximal reachable levels of internal ammonium forms}
\label{tab:results1}
\end{table}

The results have shown that $NH_3in$ does not exceed its initial level
no matter how the external ammonium is distributed between $NH_3ex$
and $NH_4^+ex$. The upper bound concentration considered for both
$NH_3ex$ and $NH4^+ex$ has been set to $1\cdot 10^{-5}$ which goes
with the typical level of concentration of the gas in the environment.

In the case of $NH_4in$ we have found that the upper bound to
maximal reachable level is in the interval $\beta\in\langle 5.3\cdot
10^{-4},5.4\cdot 10^{-4}\rangle$. Since the counterexample achieved can be a
spurious one due to the overapproximating abstraction, the exact maximal
reachable value may be lower. To this end we have conducted several
numerical simulations which give us the argument that our estimation
of $\beta$ is plausible.

\subsubsection{Dependence of stable internal ammonium on changes in external conditions}

In the second experiment, we have focused on determining how much
external ammonium have to be increased in particular form in order to
stimulate $NH3in$ to exceed the considered initial level. The setting
of partitions and initial conditions has been considered the same as
in the previous experiments.

First, we have varied the constant amount of $NH_3ex$ to find at which
level of $NH_3ex$ the maximal reachable level of $NH_3in$ is affected.
More precisely, we have observed for which setting of $NH_3ex$ the
property $\varphi\equiv\G (NH_3in < 1.1\cdot 10^{-6})$ is true and for
which it is not. The relevant experiments are summarised in
Table~\ref{tab:results2}. We have found out that if $NH_3ex$ is set to
$19.6\cdot 10^{-4}$ or higher level then $NH_3in$ increases above the
upper initial bound $1.1\cdot 10^{-6}$. The counterexamples returned
again agree with numerical simulations.

\begin{table}
{\small
\begin{center}
\begin{tabular}{|c|c|c|c||c|c|c|}
\hline
\multicolumn{4}{|c|}{Partition $(1)$} & \multicolumn{3}{|c|}{Partition $(2)$}\\
\hline
$NH_3ex$ & $\varphi$ & $\#$ states & Time ($\#$ nodes) & $\varphi$ & $\#$ states &  Time ($\#$ nodes)\\\hline
$19.5\cdot 10^{-4}$ & true & 901 & 0.22 s (1) & true & $1.4\cdot 10^5$ & 1.9 s (36)\\
$19.6\cdot 10^{-4}$ & false & 1261 & 0.6 s (1) & false & $3.4\cdot 10^5$ & 5.9 s (36)\\
\hline
\end{tabular}
\end{center}
}
\caption{Experiments on detecting $NH_3ex$ levels affecting maximal reachable $NH_3in$}
\label{tab:results2}
\end{table}

Finally, we have varied the amount of $NH_4^+ex$ in order to find at
which level of $NH_4^+ex$ the maximal reachable level of $NH_3in$ is
affected. The results presented in Table~\ref{tab:results3} give us
the conclusion that despite the level of $NH_4^+ex$ (checked up to
$10^{12}$), the maximal level reached by $NH_3in$ remains the same. In
particular, $NH_3in$ does not exceed the initial bounds.

\begin{table}
{\small
\begin{center}
\begin{tabular}{|c|c|c|c|}
\hline
\multicolumn{4}{|c|}{Partition $(1)$}\\\hline
$NH_4ex$ & $\varphi$ & $\#$ states &  Time ($\#$ nodes)\\\hline
$1$ & true & 901 & 0.25 s (1) \\
$1\cdot 10^{12}$ & true & 901 & 0.25 s (1)\\
\hline
\end{tabular}
\end{center}
}
\caption{Experiments on detecting  $NH_4^+ex$ levels affecting maximal reachable $NH_3in$}
\label{tab:results3}
\end{table}

\subsubsection{Performance}


All the experiments have been performed on a homogeneous cluster
allowing computation on up-to 22 nodes each equipped with 16GB of RAM
and a quad-core processor Intel Xeon 2GHz.  The model we have dealt
with contained $5$ dynamic variables and $2$ constants.
With partition $(1)$ the generated state space had maximally $4753$
states and with partition $(2)$ maximally $3.4\cdot 10^5$.
When trying to run one more step of automatic partition refinement,
the number of thresholds exceeded the memory reserved for storing of
the mathematical model. Note that the model has to be stored in the
memory of each node. Only the state space is distributed over the
cluster.

\section{Conclusion}


In this paper, we have presented the current version of BioDiVinE tool
which implements rectangular abstraction of continuous models of
biochemical reaction systems. The tool provides the framework for
specification and analysis of biochemical systems. The supported
analysis technique is based on the model checking method. Linear
temporal logic is used for encoding of the properties to be observed
on abstracted systems.


The tool provides parallel model checking algorithms that allow fast
response times of the analysis. We have provided a case study on which
the key features of the tool are demonstrated. The case study has
showed that the tool can be used for quickly getting the approximation
of maximal reachable concentration levels of individual species in the
model. In general, we have analysed the model with respect to the set
of safety properties which are technically tackled by construction of
the state space reachable from the given initial states. We have found
out that the main advantage of the rectangular abstraction is the
possibility to analyse the system with uncertain initial conditions.


The current drawback of the abstraction method is strong
overapproximation of non-transient states in multi-affine systems.  In
consequence, analysis of liveness kind of properties (e.g.,
oscillations, instability) is infeasible because of large amount of
spurious counterexamples that come from false identification of
non-transient states. However, this is not the case for affine systems
on which liveness properties can be checked without these problems.
Since the analysed systems are typically non-affine, we can still
employ the liveness checking on their linearised approximations.
However, by the linearisation process the precision of the analysis is
significantly reduced. Improving the tool in these aspects requires
further research.


In general, we leave for future work the development of methods for
identification of spurious paths. We think that one of the promising
directions in using the discrete abstractions for analysis of
biological models is employing the model-checking-based analysis for
extensive exploration of properties. In particular, instead of
returning only one path, the model checker should provide a set of
paths. In this directions we aim to continue the research.


\bibliographystyle{eptcs} 
\bibliography{compmod09}



\end{document}